\documentclass[twocolumn]{pasj01}

\usepackage{ulem}
\usepackage{natbib}
\usepackage{color}

\newcommand{\Dmax}{D_{\rm max}}
\newcommand{\Dmin}{D_{\rm min}}
\newcommand{\dNacc}{\dot{N}_{\rm acc}}
\newcommand{\dNaccz}{\dot{N}_{{\rm acc}, 0}}
\newcommand{\dMacc}{\dot{M}_{\rm acc}}
\newcommand{\dMaccz}{\dot{M}_{{\rm acc}, 0}}
\newcommand{\kms}{{\rm km~s^{-1}}}
\newcommand{\Macc}{M_{\rm acc}}
\newcommand{\icarus}{Icarus}
\begin{document}

\title{Metal Pollution of Low-Mass Population III Stars through
  Accretion of Interstellar Objects like `Oumuamua}

\author{
  Ataru Tanikawa\altaffilmark{1,2},
  Takeru K. Suzuki\altaffilmark{1},
  Yasuo Doi\altaffilmark{1}
}

\altaffiltext{1}{Department of Earth Science and Astronomy, College of
  Arts and Sciences, The University of Tokyo, 3-8-1 Komaba, Meguro-ku,
  Tokyo 153-8902, Japan; tanikawa@ea.c.u-tokyo.ac.jp}

\altaffiltext{2}{RIKEN Advanced Institute for Computational Science,
  7-1-26 Minatojima-minami-machi, Chuo-ku, Kobe, Hyogo 650-0047,
  Japan}

\email{tanikawa@ea.c.u-tokyo.ac.jp}


\maketitle

\begin{abstract}

We calculate accretion mass of interstellar objects (ISOs) like
`Oumuamua onto low-mass population III stars (Pop.~III survivors), and
estimate surface pollution of Pop.~III survivors. An ISO number
density estimated from the discovery of `Oumuamua is so high ($\sim
0.2$~au$^{-3}$) that Pop.~III survivors have chances at colliding with
ISOs $\gtrsim 10^5$ times per $1$~Gyr. `Oumuamua itself would be
sublimated near Pop.~III survivors, since it has small size, $\sim
100$~m. However, ISOs with size $\gtrsim 3$~km would reach the
Pop.~III survivor surfaces. Supposing an ISO cumulative number density
with size larger than $D$ is $n \propto D^{-\alpha}$, Pop.~III
survivors can accrete ISO mass $\gtrsim 10^{-16}M_\odot$, or ISO iron
mass $\gtrsim 10^{-17}M_\odot$, if $\alpha < 4$. This iron mass is
larger than the accretion mass of interstellar medium (ISM) by several
orders of magnitude. Taking into account material mixing in a
convection zone of Pop.~III survivors, we obtain their surface
pollution is typically [Fe/H] $\lesssim -8$ in most cases, however the
surface pollution of Pop.~III survivors with $0.8M_\odot$ can be
[Fe/H] $\gtrsim -6$ because of the very shallow convective layer. If
we apply to Pop.III survivors located at the Galactocentric distance
of 8 kpc, the dependence of the metal pollustion is as follows. If
$\alpha > 4$, Pop.~III survivors have no chance at colliding with ISOs
with $D \gtrsim 3$~km, and keep metal-free. If $3 < \alpha < 4$,
Pop.~III survivors would be most polluted by ISOs up to [Fe/H] $\sim
-7$. If $\alpha < 3$ up to $D \sim 10$~km, Pop.~III survivors could
hide in metal-poor stars so far discovered. Pop.~III survivors would
be more polluted with decreasing the Galactocentric distance. Although
the metal pollution depends on $\alpha$ and the Galactocentric
distance, we first show the importance of ISOs for the metal pollution
of Pop.~III survivors.

\end{abstract}

\section{Introduction}
\label{sec:introduction}

Population III (Pop.~III) stars, metal-free stars, or first stars are
epoch-making objects in the universe history. They bring an end to the
universe's dark ages, and mark the opening of metal enrichment in the
universe. It is also interesting that their formation mode is
completely different from those of Pop.~I and II stars. Their typical
mass is theoretically predicted to be $10$ -- $1000M_\odot$
\citep{1998ApJ...508..141O,2002Sci...295...93A,2004ARA&A..42...79B,2008Sci...321..669Y,2011Sci...334.1250H,2011MNRAS.413..543S,2012MNRAS.422..290S,2013RPPh...76k2901B,2013ApJ...773..185S,2014ApJ...792...32S,2015MNRAS.448..568H}. Direct
observations of Pop.~III stars are essential to investigate the
Pop.~III star era and Pop.~III stars themselves. Since massive stars
with $>10M_\odot$ have short lifetimes $\sim 10$~Myr, Pop.~III stars
should be explored in the high-redshift universe. So, the direct
observation is quite difficult, and consequently they have not yet
been detected so far. \cite{2018Natur.555...67B} have reported an
observation for a relic of Pop.~III stars, although further
confirmation is required, since the signal is much stronger than
predicted by existing cosmological models \citep{2018Natur.555...71B}.

Alternatively, Pop.~III stars can be explored in the Galaxy. If they
are born as low-mass stars, they have longer lifetimes than the Hubble
time. Low-mass Pop.~III stars are thought to be formed in the
circumstellar disk around massive Pop.~III stars
\citep{2008ApJ...677..813M,2011ApJ...727..110C,2011Sci...331.1040C,2011ApJ...737...75G,2012MNRAS.424..399G,2013MNRAS.435.3283M,2014ApJ...792...32S,2016MNRAS.463.2781C}.
We call such low-mass Pop.~III stars ``Pop.~III survivors''. However,
Pop.~III survivors have not been found, although great efforts have
been taken to \citep[e.g.][]{2006ApJ...639..897A,2015ARA&A..53..631F}.

One possibility of the absence of Pop.~III survivors is that Pop.~III
survivors suffer from metal pollution through accretion of
interstellar medium (ISM)
\citep{1981A&A....97..280Y,2015ApJ...808L..47K,2017MNRAS.469.4012S}.
\cite{2015ApJ...808L..47K} have considered Bondi-Hoyle-Lyttleton
accretion of ISM, and have asserted some metal-poor stars can be
Pop.~III survivors polluted by ISM. However,
\cite{2015MNRAS.453.2771J} have shown radiation pressure prevents
accretion of dust in ISM, and \cite{2017ApJ...844..137T} have shown
stellar wind prevents accretion of gas in ISM. Although stellar wind
in their model is Pop.~I stellar wind, \cite{2018PASJ..tmp...35S} have
made clear that stellar wind of metal-poor stars (Pop.~II and III
stars) prevents the ISM accretion more strongly than that of Pop.~I
stars. Eventually, Pop.~III survivors have iron abundance [Fe/H] only
up to $\sim -14$ \citep{2017ApJ...844..137T}. This metallicity is much
smaller than currently discovered very metal deficient stars
\citep[e.g.][]{2014Natur.506..463K}.

Recently, \cite{2017Natur.552..378M} have discovered the first
interstellar object (ISO) or interstellar asteroid, called
`Oumuamua. They have estimated the ISO number density is $\sim
0.1$~au$^{-3}$. \cite{2018ApJ...855L..10D} have also inferred the ISO
number density $\sim 0.2$~au$^{-3}$ from an estimate of the Pan-STARRS
survey volume. This number density is so high that ISOs can plunge
into and pollute Pop.~III survivors many times in lifetimes of
Pop.~III survivors. In this paper, we calculate an ISO accretion rate
onto Pop.~III survivors, and their metal pollution.

This paper is structured as follows. In
section~\ref{sec:AccretionRate}, we calculate an ISO accretion rate
onto Pop.~III survivors. In section~\ref{sec:Discussion}, we estimate
metallicity of polluted Pop.~III survivors, taking into account
surface convection zones of Pop.~III survivors. In
section~\ref{sec:Summary}, we summarize this paper.

\section{Accretion Rate}
\label{sec:AccretionRate}

We can express an ISO accretion rate onto Pop.~III survivors in number as
\begin{eqnarray}
  \dNacc = f n \sigma v, \label{eq:dnacc1}
\end{eqnarray}
where $n$ is an ISO cumulative number density with ISOs' radii larger
than $D$, $\sigma$ is cross section of collision between ISOs and
Pop.~III survivors, and $v$ is a relative speed between ISOs and
Pop.~III survivors. The value $f$ is a fraction of an ISO-rich region
in an orbit of a Pop.~III survivor. Next, we write an ISO accretion
rate in mass as
\begin{eqnarray}
  \dMacc = \int_{\Dmax}^{\Dmin} \left\{ f \frac{dn}{dD} \sigma v
  \left[ m_0 \left( \frac{D}{D_0} \right)^3 \right] \right\}
  dD, \label{eq:dmacc1}
\end{eqnarray}
where $m_0$ is the mass of an ISO with its radius $D_0$, $\Dmin$ is
the minimum radius of an ISO reaching a Pop.~III survivor surface
without sublimation, and $\Dmax$ is the maximum radius of an ISO
colliding with a Pop.~III survivor once at least. We assume the ISO
cumulative number density can be written as a single power-law
function. Then, we give the cumulative number
density as
\begin{eqnarray}
  n = n_0 \left( \frac{D}{D_0} \right)^{- \alpha}, \label{eq:ndens}
\end{eqnarray}
where $n_0$ is the ISO cumulative number density with its radius
larger than $D_0$. From the observation of `Oumuamua, we adopt $n_0
\sim 0.2$~au$^{-3}$, and $D_0 \sim 100$~m in this paper
\citep{2018ApJ...855L..10D}. Since the power $\alpha$ has
not yet been constrained strictly even from an estimate of the
Pan-STARRS survey volume \citep{2018ApJ...855L..10D}, we consider a
wide range of the power $\alpha$. Rewriting
Equation~(\ref{eq:dmacc1}), we finally obtain the following equation:
\begin{eqnarray}
&\dMacc = \dMaccz \nonumber \\
  &\times \left\{
  \begin{array}{lc}
    \displaystyle \frac{\alpha}{\alpha-3} \left[ \left(
      \frac{\Dmin}{D_0} \right)^{-\alpha+3} - \left( \frac{\Dmax}{D_0}
      \right)^{-\alpha+3} \right] & (\alpha > 3), \\
    \displaystyle \alpha \left[ \log(\Dmax) - \log(\Dmin) \right] &
    (\alpha = 3), \\
    \displaystyle \frac{\alpha}{3-\alpha} \left[ \left(
      \frac{\Dmax}{D_0} \right)^{3-\alpha} - \left( \frac{\Dmin}{D_0}
      \right)^{3-\alpha} \right] & (\alpha < 3),
  \end{array}
\right. \label{eq:dmacc2} 
\end{eqnarray}
where
\begin{eqnarray}
  \dMaccz &= m_0 \dot{N}_{\rm acc,0}, \\
  \dNaccz &= f n_0 \sigma v. \label{eq:dnacc0}
\end{eqnarray}
The right sides of Equation~(\ref{eq:dmacc2}) are the same in the
cases of $\alpha>3$ and $<3$. We divide these cases for
visibility. The total mass of ISOs can be written as
\begin{eqnarray}
  M_{\rm iso} &= \int \frac{dn}{dD} \left[ m_0 \left( \frac{D}{D_0}
    \right)^3 \right] dD \\ 
  &= - \frac{\alpha m_0 n_0}{D_0} \int \left( \frac{D}{D_0}
  \right)^{-\alpha+2} dD.
\end{eqnarray}
Note that the total mass of ISOs diverges for $\alpha \le 3$ if the
power $\alpha$ keeps constant at $D \rightarrow \infty$. When we adopt
$\alpha \le 3$, we suppose there are a knee or cutoff at some size
$D$.

Now, we calculate the accretion rate in number, $\dNaccz$. The
distribution of ISOs is concentrated in the Galactic disk region that
consists of more metal-rich Pop.~I stars, because ISOs are themselves
made from heavy elements.  Therefore, we can safely assume that ISOs
orbit around the Galaxy with the Galactic disk at a circular velocity
of the Galaxy, $\sim 220 \kms$. On the other hand, Pop.~III survivors
must have been formed before the formation of the Galactic disk. They
would wander in the Galactic halo \citep[e.g.][]{2016ApJ...826....9I},
and are distributed in an isotropic manner with the average circular
velocity, $\sim 220$~km~s$^{-1}$. Eventually, a typical relative speed
between ISOs and Pop.~III survivors would be $\sqrt{2}$ times the
circular velocity, i.e. $v \sim 310 \kms$. Pop.~III survivors would
accrete ISOs only when they traverse the Galactic disk twice an
orbit. Let us consider, as a typical example, a Pop.~III survivor that
orbits at a distance from the Galactic center with the inclination
angle of $30$ degree with respect to the Galactic plane. This
inclination angle is the average value in isotropic velocity
distribution. If we take $400$~pc for the thickness of the Galactic
disk, we obtain $f$ in equation~(\ref{eq:dnacc1}) is $\sim 0.032$. We
may underestimate $f$. Pop.~III survivors spend longer time orbiting
in an ISO-rich region with decreasing the Galactocentric distance,
since the Galactic disk becomes thicker, and the Galactic bulge is
present at the Galactic center. Note that Pop.~III survivors could be
preferentially concentrated at the Galactic center, such as the
Galactic bulge
\citep{2006ApJ...653..285S,2010MNRAS.401L...5S,2010ApJ...708.1398T}.
Considering gravitational focusing, we obtain the cross section
$\sigma$ as
\begin{eqnarray}
  \sigma &= \pi r_{*}^2 \left( 1 + \frac{2 G M_{*}}{r_{*} v^2}
  \right),
\end{eqnarray}
where $r_{*}$ and $M_{*}$ are respectively the radius and mass of a
Pop.~III survivor, and $G$ is the gravitational constant. We adopt the
solar radius and mass for $r_{*}$ and $M_{*}$, respectively. This is
because Pop.~III survivors have $\lesssim 0.8M_\odot$ and similar
$M_{*}/r_{*}$ to that of the Sun \citep{2002ApJ...580.1100R}. Then, we
obtain $\sigma \sim 7.6 \cdot 10^{22}$~cm$^{2}$. Using the above $f$,
$\sigma$, and $v$, we get $\dNaccz$ as
\begin{eqnarray}
  \dNaccz \sim 1.4 \cdot 10^{-4} \left( \frac{n_0}{0.2~\mbox{au}^{-3}}
  \right)~\mbox{[yr$^{-1}$]}.
\end{eqnarray}
As is clear from the above equation, Pop.~III survivors have chances
at accreting a large number of ISOs in their lives, $1.4 \cdot 10^5$
times per $1$~Gyr.

Before proceeding to this calculation, we show accretion rates (or
collision rates) of larger objects such as stars and planets are
extremely small. In the solar neighborhood, stellar number density is
$\sim 0.1$~pc$^{-3}$. Then, $\dNaccz \sim 8.8 \cdot
10^{-21}$~yr$^{-1}$ for stars. The number density of free floating
planets \citep{2011Natur.473..349S} could be $2000$ times higher than
the stellar number density \citep{2018ApJ...853L..27D}. Nevertheless,
the collision rate is $\sim 1.8 \cdot 10^{-17}$~yr$^{-1}$ for free
floating planets. It is clear that Pop.~III survivors have no chance
to collide with other stars and free floating planets.

We can obtain the accretion rate in mass, $\dMaccz$, as
\begin{eqnarray}
  \dMaccz \sim 9.9 \cdot 10^{-25} \left( \frac{m_0}{1.4 \cdot
    10^{13}~\mbox{g}} \right) \left( \frac{n_0}{0.2~\mbox{au}^{-3}}
  \right)~\mbox{[$M_\odot$~yr$^{-1}$]},
\end{eqnarray}
where we assume the mass density of a spherical ISO is
$3$~g~cm$^{-3}$, when we derive $m_0$ for $D_0=100$~m, which is a
typical value of asteroids \citep{2012P&SS...73...98C}.

When ISOs approach to Pop.~III survivors, they would be strongly
radiated, and completely sublimated if they are small. If they are
sublimated, their debris would be blown away by stellar wind, and
would not be accreted onto Pop.~III survivors. Here, we estimate
$\Dmin$, the minimum size of ISOs that reach Pop.~III survivors
without sublimated. An ISO with its radius $D$ spends a certain amount
of time ($\Delta t_{\rm cond}$) sublimated after it attains its
sublimation temperature on its surface. Supposing thermal energy is
conducted through diffusion process, we can give $\Delta t_{\rm cond}$
as
\begin{eqnarray}
  \Delta t_{\rm cond} \sim \frac{D^2}{\kappa} \label{eq:diffusiontime}
\end{eqnarray}
where $\kappa$ is thermal conductivity of an ISO. We express a
distance between an ISO and Pop.~III survivor when the ISO attains its
sublimation temperature on its surface under the radiative equilibrium
as follows:
\begin{eqnarray}
  &R = \left( \frac{L_{*}}{4 \pi \sigma_{\rm s}T^4} \right)^{1/2},
  \\
  &\sim 6.9 \cdot 10^{-2} \left( \frac{L_{*}}{3.8 \cdot 10^{33}
    \mbox{erg~s$^{-1}$}} \right)^{1/2} \left( \frac{T}{1500 \mbox{K}}
  \right)^{-2} \; \mbox{[au]} \label{eq:distance}
\end{eqnarray}
where $L_{*}$ is the bolometric luminosity of a Pop.~III survivor, $T$
is the sublimation temperature of an ISO, and $\sigma_{\rm s}$ is the
Stefan-Boltzmann constant. For the second equality, we adopt the solar
luminosity for $L_{*}$, and typical sublimation temperature of dust
grains \citep[e.g.][]{1994ApJ...421..640N} for $T$. We set the ISO's
albedo to zero, which is based on the albedo of `Oumuamua assumed by
\cite{2017Natur.552..378M}, $\sim 0.04$. This assumption of
  albedo increases $\Dmin$, and conservatively reduces metal pollution
  of Pop.~III survivors, although some asteroids have albedo $\sim
  0.2$ \citep{2016AJ....152...79W}. The velocity of an ISO at a
distance $R$ is calculated as
\begin{eqnarray}
  v_{\rm R} &= \left( v^2 + \frac{2GM_{*}}{R} \right)^{1/2} \\
  &\sim 3.5 \cdot 10^7 \; \mbox{[cm~s$^{-1}$]},
\end{eqnarray}
where we adopt $v=310$~km~s$^{-1}$, $M_{*}=1M_\odot$, and $R=0.069$~au
for the second equality. An ISO spends the time $R/v_{\rm R}$ reaching
Pop.~III survivors after the surface starts to be sublimated. We
equate $R/v_{\rm R}$ with $\Delta t_{\rm cond}$ for $D=\Dmin$ in
Equation~(\ref{eq:diffusiontime}). Using Equation~(\ref{eq:distance}),
we can estimate $\Dmin$, such that
\begin{eqnarray}
  &\Dmin \sim 3.0 \left( \frac{\kappa}{3 \cdot 10^6
    \mbox{erg~cm$^{-1}$~K$^{-1}$}} \right)^{1/2} \\
  &\times \left( \frac{L_{*}}{ 3.8 \cdot 10^{33} \mbox{erg~s$^{-1}$}}
  \right)^{1/4} \left( \frac{T}{1500 \mbox{K}} \right)^{-1} \;
  \mbox{[km]}.
\end{eqnarray}
We adopt the thermal conductivity of iron at $1000$~K for $\kappa$,
since `Oumuamua is rocky asteroid, not icy comet. Although hot corona
is expected to exist in the stellar atmosphere
\citep{2018PASJ..tmp...35S}, we expect that its effect is not
essential for ISOs with $D>1$~km because the density of the corona is
significantly low.

When an asteroid has size of $3.0$~km, it has $\sim 3.4 \cdot
10^{17}$~g, where the mass density is assumed to $3$~g~cm$^{-3}$. On
the other hand, comets with size of $\sim 10^{18}$~g can reach the
solar photosphere \citep{2015ApJ...807..165B}. Our $\Dmin$ could be
consistent with the minimum size of comets plunging into the Sun,
since comets are volatile whereas asteroids are not.

We derive $\Dmax$, the maximum radius of ISOs colliding with Pop.~III
survivors once at least. The number density of ISOs increases with
time via metal enrichment in the Galaxy. As a result, the ISO
cumulative number density is expected to be comparable to the present
value in the last $\sim$ few Gyr; we here define $\Delta t_{\rm iso}$
for this duration. Then we can derive $D=D_{\rm max}$ from $\dNacc
\Delta t_{\rm iso} \sim 1$. Using Equation~(\ref{eq:dnacc1}),
(\ref{eq:ndens}), and (\ref{eq:dnacc0}), we can write $\Dmax$ as
\begin{eqnarray}
  \Dmax \sim D_0 \left(\dNaccz \Delta t_{\rm iso} \right)^{1/\alpha}.
\end{eqnarray}

\begin{figure}[ht!]
  \includegraphics[width=8cm]{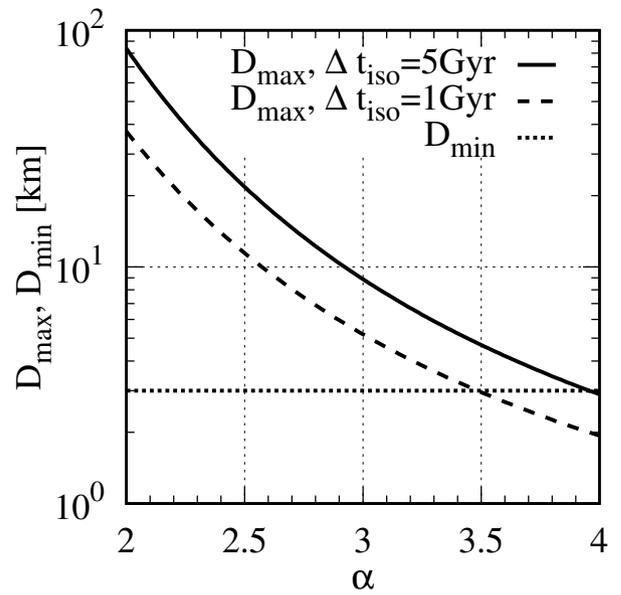}
  \caption{Maximum and minimum size of ISOs which Pop.~III survivors
    can accrete ($\Dmax$ and $\Dmin$, respectively) as a function of
    the power $\alpha$. The solid and dashed curves indicate $\Dmax$
    for $\Delta t_{\rm iso}=5$~Gyr and $1$~Gyr, respectively. The
    dotted curve indicates $\Dmin$. \label{fig:dmax}}
\end{figure}

The actual value of $\Delta t_{\rm iso}$ is unknown. So, we assume
$\Delta t_{\rm iso} \sim 5$~Gyr and $1$~Gyr. The former ($\Delta
t_{\rm iso} \sim 5$~Gyr) is equivalent to the solar age, or to the age
of the Galactic disk at the solar neighborhood
\citep[e.g.][]{2017MNRAS.472.3637G}. ISOs would be formed
simultaneously with the Galactic disk formation, if they are ejected
from the inner protoplanetary disk
\citep{2017RNAAS...1...13G,2017arXiv171103558P}. ISOs would be formed
$< 1$~Gyr after the Galactic disk formation, if their progenitors are
a sort of the Oort cloud around intermediate-mass stars with
$2-8M_\odot$, and are released when the intermediate-mass stars enter
into asymptotic giant branch phases
\citep{2011MNRAS.417.2104V}. Regardless of the formation scenarios of
ISOs, ISOs could be in the Galactic disk for $\Delta t_{\rm iso} \sim
5$~Gyr. We adopt the latter ($\Delta t_{\rm iso} \sim 1$~Gyr) in order
to take into account timescale on which ISOs accumulate in the
Galactic disk for more conservative constraints. Figure~\ref{fig:dmax}
shows $\Dmax$ as well as $\Dmin$ as a reference.

\begin{figure}[ht!]
  \includegraphics[width=8cm]{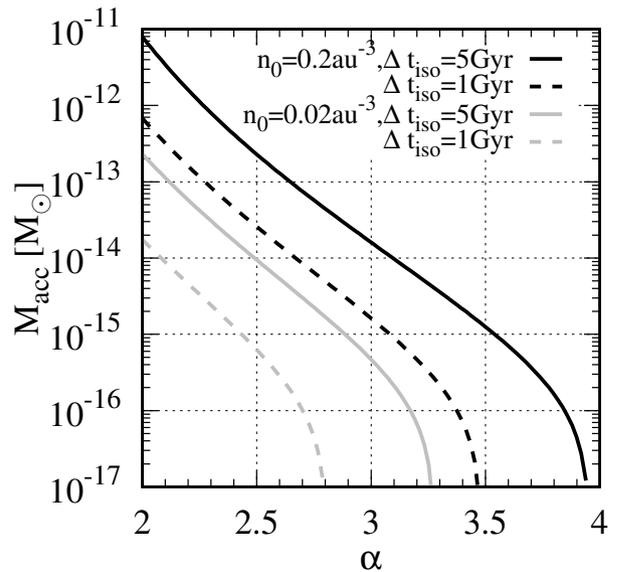}
  \caption{Total accretion mass as a function of the power $\alpha$.
    Black and gray curves show the cases of $n_0=0.2$ and
    $0.02$~au$^{-3}$.  Solid and dashed curves indicate $\Delta t_{\rm
      iso}=5$~Gyr and $1$~Gyr, respectively. \label{fig:macc}}
\end{figure}

We calculate the total accretion mass of ISOs onto a Pop.~III
survivor, $\Macc \sim \dMacc \Delta t_{\rm iso}$, using
Equation~(\ref{eq:dmacc2}). We draw $\Macc$ as a function of $\alpha$
in Figure~\ref{fig:macc}. Figure~\ref{fig:macc} shows steep decrease
of $\Macc$ at $\alpha \sim 4$ for $\Delta t_{\rm iso}=5$~Gyr and at
$\alpha \sim 3.5$ for $\Delta t_{\rm iso}=1$~Gyr due to $\Dmax <
\Dmin$ (see Figure~\ref{fig:dmax}) for $n_0=0.2$~au$^{-3}$.  We can
see $\Macc \gtrsim 10^{-16}M_\odot$, unless $\Dmax < \Dmin$, or
$\alpha$ is large. On the analogy of the cumulative number density of
asteroids in the main belt, those in Edgeworth-Kuiper belt, and long
period comets from sub-km to km
\citep[][respectively]{2009Icar..202..104G,2004AJ....128.1916K,2012MNRAS.423.1674F},
the power $\alpha$ could be close to, or shallower than $3$.

Our $\Macc$ is even larger than ISM's $\Macc$ by several orders of
magnitude. ISOs would contain about $10$~\% iron in mass, similarly to
the solar compositions \citep{2009ARA&A..47..481A}. Thus, Pop.~III
survivors accrete iron mass of $\gtrsim 10^{-17}M_\odot$ through
collision with ISOs. On the other hand, \cite{2017ApJ...844..137T}
have shown the total accreted iron mass from the gas component of ISM
is $\lesssim 10^{-19}M_\odot$ from ISM accretion.

The estimated value of $n_0$ can contain large uncertainties, since
`Oumuamua is only one ISO so far discovered.  We pessimistically
decrease $n_0$ from $0.2$~au$^{-3}$ to $0.02$~au$^{-3}$ in order to
examine an effect of $n_0$ on metallicity of polluted Pop.~III
survivors. Figure~\ref{fig:macc} also shows $\Macc$ of Pop.~III
survivors when $n_0=0.02$~au$^{-3}$. The $\Macc$ decreases by more
than an order of magnitude. This is because $\Dmax$ as well as
$\dNaccz$ becomes smaller with $n_0$ decreasing. Moreover, $\Macc$
steeply decreases at smaller $\alpha$ than in the case of $n_0 =
0.2$~au$^{-3}$ due to smaller $\Dmax$. Nevertheless, $\Macc \gtrsim
10^{-16}M_\odot$, if $\alpha \lesssim 3$. In conclusion, ISOs can be
the most dominant polluters of Pop.~III survivors.

\section{Discussion}
\label{sec:Discussion}

We estimate surface pollution of Pop.~III survivors, considering the
thickness of their convection zones under their surfaces. Accreting
metals are mixed only within the surface convective zone and do not
leak downward into the stable radiative zone. According to
\cite{2002ApJ...580.1100R}, metal-poor stars with $\lesssim
0.8M_\odot$ have their lifetimes $> 12$~Gyr. So, we suppose Pop.~III
survivors that were born after $< 1$~Gyr of the Big Bang and have mass
of $\lesssim 0.8M_\odot$. In the cases of $0.75M_\odot$ and
$0.7M_\odot$ stars, the mass fractions of convection zones are
respectively $10^{-2.5}$ and $10^{-2}$ in the last $5$~Gyr. On the
other hand, in a $0.8M_\odot$ star the mass fraction of a convection
zone rapidly decreases with time from $10^{-3.5}$ at $\approx 5$~Gyr
ago and $10^{-6}$ at $\approx 1$~Gyr ago.

We calculate metallicity of a Pop.~III survivor as follow:
\begin{eqnarray}
  \mbox{[Fe/H]} \sim \log_{10} \left( \frac{1}{f_{\rm conv}}
  \frac{\dMacc \Delta t_{\rm pol}}{M_{*} Z_\odot} \right).
\end{eqnarray}
We set the mass fraction of metals in the Sun, $Z_\odot$, to $1.4$~\%
\citep{2009ARA&A..47..481A}. We set the mass fraction of a surface
convection zone, $f_{\rm conv}$, in reference to
\cite{2002ApJ...580.1100R} as follows. For $M_{*}=0.7$ and
$0.75M_\odot$, we adopt $f_{\rm conv}=10^{-2}$ and $10^{-2.5}$,
respectively, and $\Delta t_{\rm pol} = \Delta t_{\rm iso}$. For
$M_{*}=0.8M_\odot$ with $\Delta t_{\rm iso} = 1$~Gyr, we adopt $f_{\rm
  conv}=10^{-6}$, and $\Delta t_{\rm pol} = \Delta t_{\rm iso}$. For
$M_{*}=0.8M_\odot$ with $\Delta t_{\rm iso} = 5$~Gyr, we calculate
[Fe/H], taking into account the time dependence of the mass fraction
of a convection zone. If the Pop.~III survivor is dominantly polluted
in the last $1$~Gyr, we adopt [Fe/H] the same as that in the case of
$M_{*}=0.8M_\odot$ with $\Delta t_{\rm iso} = 1$~Gyr. If not, we
calculate [Fe/H], adopting $f_{\rm conv}=10^{-3.5}$ and $\Delta t_{\rm
  pol} = \Delta t_{\rm iso}$.

\begin{figure}[ht!]
  \includegraphics[width=8cm]{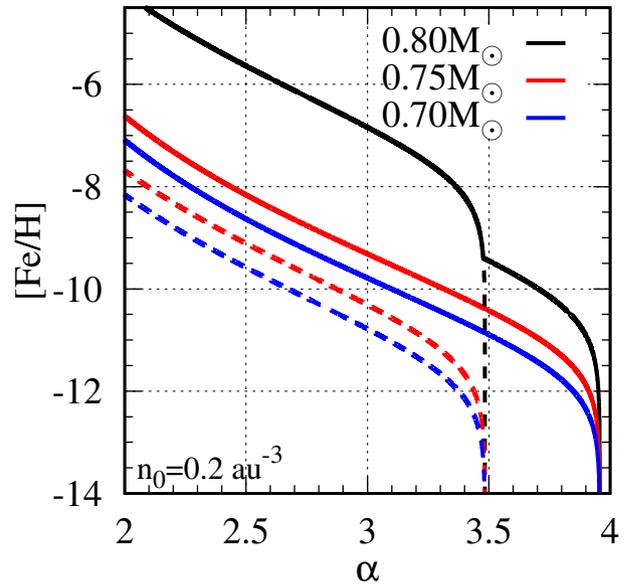}
  \caption{Metallicity of Pop.~III survivors as a function of the
    power $\alpha$. Black, red, and blue curves indicate Pop.~III
    survivors with $0.8$, $0.75$, and $0.7M_\odot$,
    respectively. Solid and dashed curves show metallicity in the
    cases of $\Delta t_{\rm iso} = 5$~Gyr and $1$~Gyr,
    respectively. For the $0.8M_\odot$ case, the solid and dashed
    curves overlap, when $\alpha<3.5$. \label{fig:pol}}
\end{figure}

We summarize the surface pollution of Pop.~III survivors in
Figure~\ref{fig:pol}, where we set $n_0 = 0.2$~au$^{-3}$. Since we
suppose ISO compositions are the same as the metal compositions of the
Sun, [Fe/H] is the same as metallicity [M/H]. For Pop.~III survivors
with $0.8M_\odot$, the metallicity in $\Delta t_{\rm iso}=5$~Gyr is
the same as in $\Delta t_{\rm iso}=1$~Gyr when $\alpha \lesssim 3.5$,
since the metal pollution in the last $1$~Gyr is dominant.  Pop.~III
survivors with $0.7$ and $0.75M_\odot$ get metallicity with [Fe/H]
$\sim -8$ at most even if $\alpha \sim 2.5$. On the other hand,
Pop.~III survivors with $0.8M_\odot$ can get metallicity with [Fe/H]
$\gtrsim -6$, if $\alpha \gtrsim 2.5$. The metallicity [Fe/H] steeply
decreases at $\alpha \sim 4$ for $\Delta t_{\rm iso} = 5$~Gyr and
$\alpha \sim 3.5$ for $\Delta t_{\rm iso} = 1$~Gyr, since $\Dmax <
\Dmin$.

We use SAGA database \citep[e.g.][]{2008PASJ...60.1159S}, and search
for metal-poor stars with [Fe/H] $<-5$. Additionally, we investigate
their effective temperature in order to conjecture their
mass. According to \cite{2002ApJ...580.1100R}, mass of a Pop.~III
survivor is $\sim 0.8M_\odot$ if its effective temperature is $>
6000$~K, and is $\lesssim 0.75M_\odot$ if not. Then, we find three
stars with [Fe/H] $< -5$: SMSS~J031300.36-670839.3 with [Fe/H] $<
-7.3$ and $\sim 5100$~K \citep{2014Natur.506..463K}, SDSS~J1035+0641
with [Fe/H] $< -5.07$ and $\sim 6300$~K \citep{2015A&A...579A..28B},
and SDSS~J131326.89-001941.4 with [Fe/H] $\sim -5.00$ and $\sim
5200$~K \citep{2015ApJ...810L..27F}. SMSS~J031300.36-670839.3 and
SDSS~J131326.89-001941.4 could be Pop.~III survivors with $\sim
0.75M_\odot$, if $\alpha < 2$ for $D \gtrsim 10^2$~km. SDSS~J1035+0641
could be a Pop.~III survivors with $\sim 0.8M_\odot$, if $\alpha
\gtrsim 2.5$ up to $D \sim 10$~km. Therefore, SDSS~J1035+0641 has the
most loose conditions of ISOs among the three metal-poor stars to be a
Pop.~III survivor.

\section{Summary}
\label{sec:Summary}

We calculated the total accretion mass of ISOs onto Pop.~III
survivors. The mass is $\gtrsim 10^{-16}M_\odot$, if the power of the
ISO cumulative number density $\alpha$ is $\lesssim 4$. We can convert
the accretion mass to iron mass $\gtrsim 10^{-17}M_\odot$. This
accretion mass is even larger than ISM accretion mass by several
orders of magnitude. Therefore, ISOs can be the most dominant
polluters of Pop.~III survivors.

We estimated the surface metallicity of Pop.~III survivors polluted by
ISOs, considering convection zones of Pop.~III survivors. If Pop.~III
survivors have $0.7M_\odot$ and $0.75M_\odot$, their metallicity can
be [Fe/H] $\lesssim -8$. On the other hand, if Pop.~III survivors have
$0.8M_\odot$, their metallicity can be enhanced to [Fe/H] $\gtrsim
-6$. This is because the mass fraction of a convection zone is down to
$10^{-6}$ when their ages are $> 10$~Gyr.

The star SDSS~J1035+0641 has metallicity of [Fe/H] $\sim -5$, and
effective temperature of $6300$~K. It can have a thin convection zone,
and could be a Pop.~III survivors, if the ISO cumulative number
density has shallow power law with $\alpha \gtrsim 2.5$ up to $D \sim
10$~km. In order to conclude whether SDSS~J1035+0641 and other
metal-poor stars are Pop.~III survivors or not, we need ISO cumulative
number density up to $D \sim 10$~km.

We note that the ISO accretion mass strongly depends on the power of
the ISO cumulative number density, $\alpha$. If we apply to Pop.III
survivors located at the Galactocentric distance of 8 kpc, the
dependence of the metal pollustion is as follows. If $\alpha>4$,
Pop.~III survivors are never polluted by ISOs. If $3 < \alpha < 4$,
Pop.~III survivors can be polluted up to [Fe/H] $\sim -7$. If
$\alpha<3$ up to $D \sim 10$~km, Pop.~III survivors could hide in
metal-poor stars so far discovered. We expect the ISO cumulative
number density will be determined in near future.

Since Pop.~III survivors could be preferentially concentrated at the
Galactic center
\citep{2006ApJ...653..285S,2010MNRAS.401L...5S,2010ApJ...708.1398T},
we may underestimate the metal pollution of Pop.~III survivors. This
is because Pop.~III survivors spend longer time orbiting in an
ISO-rich region with the Galactocentric distance decreasing. Note that
the Galactic disk becomes thicker with the Galactocentric distance,
and the Galactic bulge is present at the Galactic center. In other
words, $f$ becomes larger as the Galactocentric distance becomes
smaller.

We should derive chemical abundance of Pop.~III survivors in order to
observationally confirm that Pop.~III survivors are most polluted by
ISOs, although we discuss only about [Fe/H] in this paper. The
chemical abundance would be determined by a combination of ISO
composition and volatility. In future work, we will obtain the
chemical abundance of Pop.~III survivors polluted by ISOs.

\ack

A. Tanikawa thanks I. Hachisu and K. Kakiuchi for fruitful
discussions. This research has been supported in part by MEXT program
for the Development and Improvement for the Next Generation Ultra
High-Speed Computer System under its Subsidies for Operating the
Specific Advanced Large Research Facilities, and by Grants-in-Aid for
Scientific Research (16K17656, 17H01105, 17H06360,
18H01250) from the Japan Society for the Promotion of
Science.

%\bibliography{natbib}

\end{document}